# Goos-Hänchen shift of a spin-wave beam at the interface between two ferromagnets.

Marina Mailyan, Pawel Gruszecki, *Member, IEEE,* Oksana Gorobets, Maciej Krawczyk, *Member, IEEE*

*Abstract*—Spin waves are promising information carriers which can be used in modern magnonic devices, characterized by higher performance and lower energy consumption than presently used electronic circuits. However, before practical application of spin waves, the efficient control over spin wave amplitude and phase needs to be developed. We analyze analytically reflection and refraction of the spin waves at the interface between two ferromagnetic materials. In the model we consider the system consisting of two semi-infinite ferromagnetic media, separated by the ultra-narrow interface region with the magnetic anisotropy. We have found the Goos-Hänchen shift for spin waves in transmission and reflection, and performed detailed investigations of its dependence on the anisotropy at the interface and materials surrounding the interface. We have demonstrated possibility of obtaining Goos-Hänchen shift of several wavelengths in reflection for realistic material parameters. That proves the possibility for change of the spin waves phase in ferromagnetic materials at subwavelength distances, which can be regarded as a metasurface for magnonics.

## I. Introduction

The main object of investigation in magnonics[Kra14, Chu14, Kru10, Dem13] are spin waves (SWs) that are magnetization excitations propagating without charge transport. This excludes Joule heating and implies that application of SWs in computing devices could significantly reduce energy consumption with respect to the charge-based alternatives [Nik13, Chu15]. Furthermore, the use of SWs as information carriers would pave the way to wave-based non-Boolean computing [Csa14], holographic memories [Koz15, Ger16] or the physical realization of neural networks [Loc14]. Crucial in these applications is the manipulation of both the amplitude and the phase of SWs. In magnonics this can be realized by means of magnonic crystals or SW scattering by disturbances such as holes [Ger16], domain walls [Mac10] or magnetic elements placed on the waveguide cross junctions [Koz15]. Another possible approach, which has emerged in modern photonics, bases on the use of structuralised surfaces, referred to metasurfaces, which can be used to wave manipulation at sub-wavelength distances [Yu14].

The subject of magnonic metasurfaces is related to the investigation of the SWs amplitude and phase changes (in relation to the incident SWs) that the reflected and transmitted SWs undergo at the interface in order to obtain required functionality. Interestingly, the phase shift can also result in the occurrence of the Goos-Hänchen (GH) effect – a lateral shift along the interface between incident and reflected (or refracted) wave beam spots[Han47]. The GH effect has been already confirmed by micromagnetic simulations for exchange-dipolar SWs reflected at the edge of the magnetic thin film [Gru14]. The magnetic properties at the film edge have been shown to be important for the lateral shift of the beam [Gru15]. Moreover, recently we have theoretically demonstrated existence of the GH shift in transmission for exchange dominated SWs [Gru17]. This effect can occur during transmission of SWs through a few-nanometres wide line with modulated value of the uniaxial anisotropy in the uniformly magnetized permalloy (Py) thin film. However, the GH effect at the interface between two different ferromagnetic materials hasn't yet been studied.

In this paper, we analyse theoretically SWs reflection and refraction at the ultra-narrow interface between two ferromagnetic materials. We consider the system consisting of two semi-infinite ferromagnetic media separated by the interface region. The interface area has a different value of the uniaxial anisotropy ($K_{12}$) with respect to the extended parts of the materials. We have derived analytical formulas describing the reflectance and transmittance of the exchange SWs, as well as the phase and GH shifts for refracted and reflected SWs. In derivation we took into account: non-uniform exchange and uniaxial magnetic anisotropy of the ferromagnetic materials; the interlayer exchange coupling at the interface between the ferromagnets and the magnetic anisotropy of the interface. We show that the reflection and transmission of SWs are sensitive to slight changes in the magnetic anisotropy introduced at the interface, the thickness of which is much smaller than the wavelength of SWs. Moreover, we report

This research has received founding from National Science Centre of Poland project No. UMO-2012/07/E/ST3/00538 and from the European Union Horizon2020 research and innovation programme under the Marie Sklodowska-Curie grant agreement No. 644348 (MagIC). The computations were partially performed at Poznan Supercomputing and Networking Center (grant No 209).

M. M. Author is with National Technical University of Ukraine "Igor Sikorsky Kyiv Polytechnic Institute", 37 Peremogy ave., 03056, Kyiv, Ukraine (e-mail: marina92mmd@gmail.com).

P. G. Author is with Faculty of Physics, Adam Mickiewicz University in Poznan, Umultowska 85, Poznań, 61-614, Poland (e-mail: pawel.gruszecki@amu.edu.pl).

O. G. Author is with National Technical University of Ukraine "Igor Sikorsky Kyiv Polytechnic Institute", 37 Peremogy ave., 03056, Kyiv, Ukraine and 3Institute of Magnetism, National Academy of Sciences of Ukraine, 36-b Vernadskogo st., 03142, Kyiv, Ukraine (e-mail: pitbm@ukr.net).

M. K. Author is with Faculty of Physics, Adam Mickiewicz University in Poznan, Umultowska 85, Poznań, 61-614, Poland (e-mail: krawczyk@amu.edu.pl).

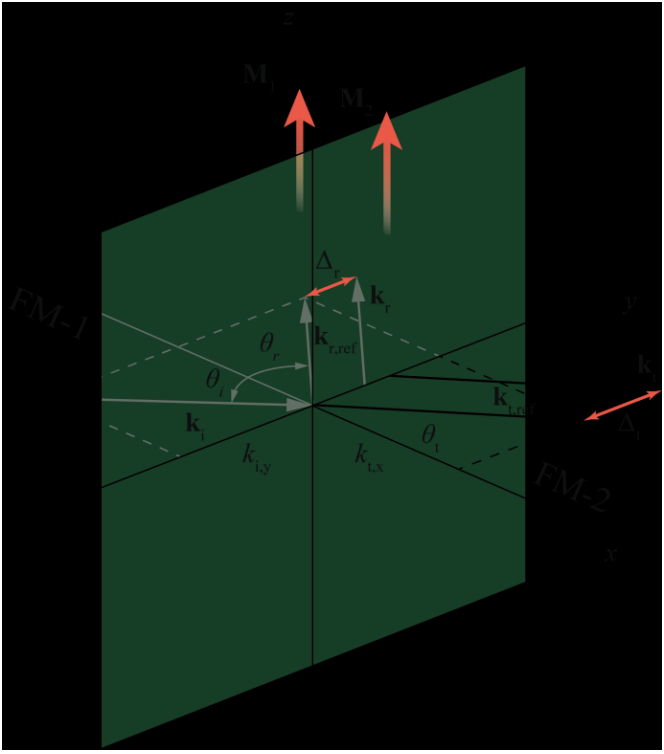

Fig. 1. Schema of the system. The structure consists of two semi-infinite ferromagnetic materials, FM-1 and FM-2, separated by the interface region (green plane) with a width $\delta$; $k_i$, $k_r$ and $k_t$ are the wavevectors of incident, reflected and transmitted SW beams, respectively; the wavevectors of reference reflected and refracted beams (with assumed zero shift) are denoted as $k_{r,\text{ref}}$ and $k_{t,\text{ref}}$, respectively; $\Delta_r$ and $\Delta_t$ is the Goos-Hänchen shift of the reflected and transmitted beam with respect to the incident beam. Angles of incidence, reflection and transmission are denoted as $\theta_i$, $\theta_r$ and $\theta_t$, respectively.

the GH shift for SWs in transmission and reflection in dependence on anisotropy at interface; and magnetization and exchange constant of the surrounding materials.

The demonstrated lateral shift of SWs refracted by an interface with a width much smaller than the wave-length of the considered waves points at the possibility of steering of SWs at subwavelength distances. Further investigation can lead to the development of more effective methods of phase modulation at sub-wavelength distances with already available technological processes. This sets a promising direction in the study of magnonic metasurfaces, a novel field in magnonics, next to the graded index magnonics [Dav15].

This paper is organized as follows. In Sec. II we present the analytical model of SW reflection and refraction. In Sec. III we describe the obtained analytical results. Conclusions can be found in Sec. IV.

## II. ANALYTICAL MODEL

In the analytical model we assume SWs propagating through the interface separating two semi-infinite ferromagnetic materials, FM-1 and FM-2 (Fig. 1). The uniform magnetization of the system is directed along the external magnetic field $\mathbf{H} = [0,0,H_0]$ parallel to the interface ($yOz$ plane) and perpendicular to the plane of the SW incidence ($xOy$ plane). In both ferromagnets we consider magnetocrystalline (uniaxial) anisotropy directed parallel to that of the surface magnetic anisotropy, along the $z$ axis: $\mathbf{n}^{(1)} = \mathbf{n}^{(2)} = \mathbf{e}_z$. We limit analysis only to the exchange SWs (high-frequency short-wavelength SWs) which are uniform along the $z$ axis.

Let us write the total energy of the system of two semi-infinite ferromagnetic media (indicated with superscripts (1) and (2)) in the uniform constant external magnetic field $H_0$

$$W = \int_V dv \Big[ w_H^{(1)} + w_H^{(2)} + w_\text{ex}^{(1)} + w_\text{ex}^{(2)} + \\ + w_\text{anis}^{(2)} + w_\text{ex}^{(12)} + w_\text{anis}^{(12)} \Big], \quad (1)$$

where the integration is evaluated over the whole volume. The term $w_H^{(l)} = -\mathbf{M}^{(l)} \cdot \mathbf{H}$ is the Zeeman energy density, where $\mathbf{M}^{(l)} = \left[ m_x^{(l)}, m_y^{(l)}, M_l \right]$ denotes the magnetization vector in the $l$-th ferromagnet and $M_l$ is its saturation magnetization. The next two terms, $w_\text{ex}^{(l)} = \frac{1}{2}\alpha_l(x)\left(\nabla \mathbf{M}^{(l)}\right)^2$, are the exchange energy densities; $\alpha_l = A_l / M_l^2$ denotes the exchange length, and $A_l$ is the exchange constant. The term $w_\text{anis}^{(l)} = -\frac{1}{2}\beta_l(x)\left(\mathbf{M}^{(l)} \cdot \mathbf{n}^{(l)}\right)^2$ is the magnetic anisotropy energy density in the $l$-th ferromagnet; $\beta_l = K_l / M_l^2$, where $K_l$ is the uniaxial magnetic anisotropy constant.

The term $w_\text{ex}^{(12)} = A_{12} \mathbf{M}^{(1)} \cdot \mathbf{M}^{(2)} \Theta_H(x) \Theta_H(-x+\delta)$ is the energy density of the interlayer exchange coupling at the interface between the ferromagnets. $\Theta_H(x)$ is the Heaviside step function, $\delta$ is the width of the interface, and $A_{12}$ is a parameter of interlayer exchange coupling constant: $A_{12} = \xi A_{\text{int},S} / (M_\text{int}^2 \delta)$, where $A_{\text{int},S}$ denotes the effective surface exchange constant of the interface ($A_{\text{int},S} = A_\text{int} / \delta$, with $A_\text{int}$ being the exchange constant of the interface region), $M_\text{int}$ is the saturation magnetization of the interface, and $\xi$ is a proportionality coefficient (we assume $\xi = 400$, to obtain perfect match in the limit of the homogeneous material filling the whole space).

The energy density of the surface magnetic anisotropy at the interface expressed in the last term in Eq. (1) has the form:
$w_\text{anis}^{(12)} = -\frac{1}{2}\Big[\beta_{12}(x)\left(\mathbf{M}^{(1)} \cdot \mathbf{n}^{(1)}\right)\left(\mathbf{M}^{(2)} \cdot \mathbf{n}^{(2)}\right)\Big]\Theta_H(x)\Theta_H(-x+\delta)$; where $\beta_{12} = K_{12} / M_\text{int}^2$ is an anisotropy parameter, with $K_{12}$ denoting the uniaxial magnetic anisotropy constant at the interface.

Using the Landau-Lifshitz (LL) equations we describe SW dynamics for both ferromagnetic materials:

$$\frac{\partial \mathbf{M}^{(l)}}{\partial t} = |\gamma|\left(\mathbf{M}^{(l)} \times \mathbf{H}_{\text{eff}}^{(l)}\right), \quad (2)$$

where $\gamma$ is the gyromagnetic ratio. Taking into account that in the linear approximation the dynamic components of the magnetization are much smaller than the saturation magnetization, $m_{x,y}^{(l)} \ll M_l$, the latter can be treated as a constant. The effective magnetic field $\mathbf{H}_{\text{eff}}^{(l)}$ in each material can be determined as the functional derivative of the total energy, defined in Eq. (1), with respect to the magnetization vector [Res041, Gor14]:

$$\mathbf{H}_{\text{eff}}^{(l)} = -\frac{\delta W}{\delta \mathbf{M}^{(l)}} = -\frac{\partial w}{\partial \mathbf{M}^{(l)}} + \sum_{\zeta \in \{x,y,z\}} \frac{d}{d\zeta} \frac{\partial w}{\partial\left(\frac{d\mathbf{M}^{(l)}}{d\zeta}\right)}, \quad (3)$$

where $w$ is the energy density, i.e., the integral kernel in Eq. (1).

SW dispersion relation in each ferromagnet is obtained taking the plane-wave solutions of the LL equations (2), $m_{x,y}^{(l)} \propto \exp\left[i(\mathbf{k}_l \cdot \mathbf{r} - \omega_l t)\right]$:

$$\omega_l(\mathbf{k}_l) = |\gamma|\left(H_0 + \beta_l M_l + M_l \alpha_l k_l^2\right) \quad (4)$$

where $\mathbf{k}_l$ is the wave vector and $\omega_l$ is the angular frequency of SWs in the $l$-th ferromagnet [Res042].

In the investigated system we describe the circularly polarized exchange SWs as monochromatic plane waves:

$$\begin{aligned} m_x^{(1)} + im_y^{(1)} &= e^{i(\mathbf{k}_i \cdot \mathbf{r} - \omega t)} + re^{i(\mathbf{k}_r \cdot \mathbf{r} - \omega t + \varphi_r)} \\ m_x^{(2)} + im_y^{(2)} &= te^{i(\mathbf{k}_t \cdot \mathbf{r} - \omega t + \varphi_t)}, \end{aligned} \quad (5)$$

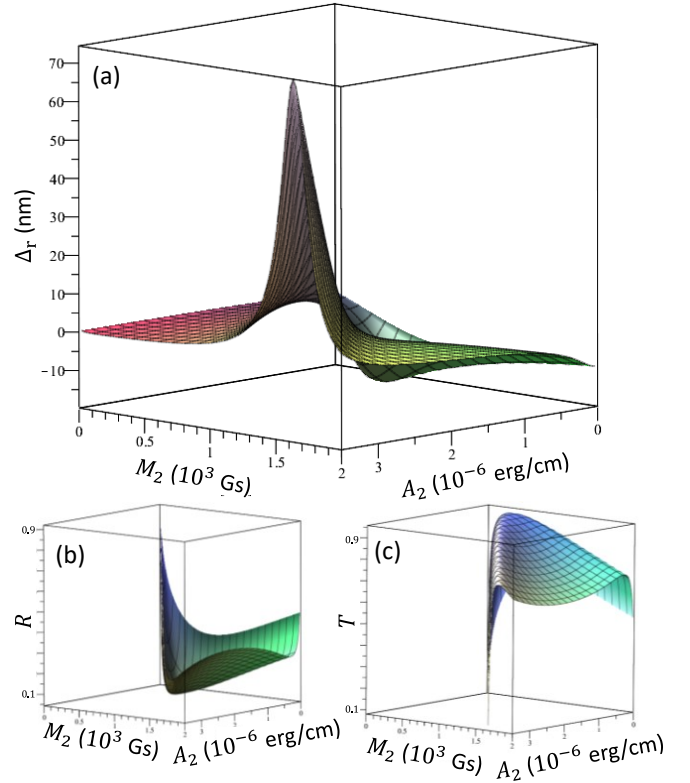

Fig. 2. (a) GH shift in reflection, (b) reflectance, and (c) transmittance in dependence on the magnetic parameters of the second medium (saturation magnetization and exchange constant). All dependencies were obtained for the interface of Py magnetic properties (exchange constant and saturation magnetization) and interfacial anisotropy of $K_{12} = 2.5 \times 10^6$ erg/cm$^3$. SWs incidence from Py under the angle of incidence equal 60° has been assumed.

where $r$ and $t$ are the reflection and transmission coefficients, respectively. The wave vectors of the incident, reflected and transmitted waves are $\mathbf{k}_i = (k_{i,x}\mathbf{e}_x + k_{i,y}\mathbf{e}_y)$, $\mathbf{k}_r = (k_{r,x}\mathbf{e}_x + k_{r,y}\mathbf{e}_y)$, and $\mathbf{k}_t = (k_{t,x}\mathbf{e}_x + k_{t,y}\mathbf{e}_y)$, respectively [Gru17]. \varphi_r and \varphi_t are phase shifts acquired by the reflected and transmitted wave at the interface with respect of the incident wave.

The boundary conditions for the dynamic components of the magnetization on both sides of the interface have been obtained by integrating the LL equations (2) with the effective magnetic field (3) over the interface region [Gru17]:

$$\begin{cases} \left(\delta A_{12} m_x^{(2)} + D m_x^{(1)} + \alpha_1 \frac{\partial m_x^{(1)}}{\partial x}\right)\Big|_{x=-0} = 0 \\ \left(\delta A_{12} m_y^{(2)} + D m_y^{(1)} + \alpha_1 \frac{\partial m_y^{(1)}}{\partial x}\right)\Big|_{x=-0} = 0 \\ \left(\delta A_{12} m_x^{(1)} + C m_x^{(2)} - \alpha_2 \frac{\partial m_x^{(2)}}{\partial x}\right)\Big|_{x=-0} = 0 \\ \left(\delta A_{12} m_y^{(1)} + C m_y^{(2)} - \alpha_2 \frac{\partial m_y^{(2)}}{\partial x}\right)\Big|_{x=-0} = 0 \end{cases} \quad (6)$$

where coefficients $C$ and $D$ are defined as follows $C = -\left[(A_{12} - B_{12})/\eta + \beta_2\right]\delta$ and $D = \left[\beta_1 - (A_{12} - B_{12})\eta\right]\delta$, where $\eta = M_2/M_1$; $-0$ and $+0$ denotes one-sided limits on the respective sides of the interface.

As a result, we receive reflectance ($R = r^2$) and transmittance ($T = t^2 \cos\theta_t / \cos\theta_i$ where $\theta_i = \arctan(k_{i,y}/k_{i,x})$ is the angle of incidence and $\theta_t = \arctan(k_{t,y}/k_{t,x})$ is the angle of refraction):

TABLE I
MATERIAL PARAMETERS

| Material | $M_S$ (Gs) | $A_{ex}$ (erg/cm) |
|---|---|---|
| Cobalt (Co) | $1.45 \times 10^3$ | $3 \times 10^{-6}$ |
| Nickel (Ni) | $0.493 \times 10^3$ | $0.85 \times 10^{-6}$ |
| Iron (Fe) | $1.74 \times 10^3$ | $2.07 \times 10^{-6}$ |
| Ni$_{0.2}$Fe$_{0.8}$ (Py) | $0.7 \times 10^3$ | $1.1 \times 10^{-6}$ |

$$R = \frac{\left(\delta^2 A_{12}^2 - CD - \alpha_1\alpha_2 k_{r,x} k_{t,x}\right)^2 + \left(D\alpha_2 k_{t,x} - C\alpha_1 k_{r,x}\right)^2}{\left(\delta^2 A_{12}^2 - CD + \alpha_1\alpha_2 k_{r,x} k_{t,x}\right)^2 + \left(D\alpha_2 k_{t,x} + C\alpha_1 k_{r,x}\right)^2}, \quad (7)$$

$$T = \frac{4\delta^2 A_{12}^2 \alpha_1\alpha_2 k_{r,x} k_{t,x}}{\left(\delta^2 A_{12}^2 - CD + \alpha_1\alpha_2 k_{r,x} k_{t,x}\right)^2 + \left(D\alpha_2 k_{t,x} + C\alpha_1 k_{r,x}\right)^2}, \quad (8)$$

the phase shifts of reflected $(\varphi_r)$ and transmitted $(\varphi_t)$ SWs is given by:

$$\varphi_r = \operatorname{arcctg}\left(\frac{D\alpha_2 k_{t,x} + C\alpha_1 k_{r,x}}{CD - \delta^2 A_{12}^2 - \alpha_1\alpha_2 k_{r,x} k_{t,x}}\right) +$$
$$\operatorname{g}\left(\frac{D\alpha_2 k_{t,x} - C\alpha_1 k_{r,x}}{\delta^2 A_{12}^2 - CD - \alpha_1\alpha_2 k_{r,x} k_{t,x}}\right), \quad (9)$$

$$\varphi_t = \operatorname{arcctg}\left(\frac{D\alpha_2 k_{t,x} + C\alpha_1 k_{r,x}}{CD - \delta^2 A_{12}^2 - \alpha_1\alpha_2 k_{r,x} k_{t,x}}\right). \quad (10)$$

Substituting to Eqs. (9)-(10) the reflected and transmitted wave vectors normal components $k_{r(t),x} = \left(k_{r(t)}^2 - k_{r(t),y}^2\right)^{1/2}$ and taking into account the equality of the tangential components of the wave vectors $k_{r,y} = k_{t,y}$, we can evaluate the GH shifts for reflected and transmitted SWs according to the stationary phase method [Dad12, Pur86]:

$$\Delta_{r(t)} = -\frac{\partial \varphi_{r(t)}}{\partial k_{r(t),y}}. \quad (11)$$

The final formulas can be found in Appendix A from Ref. [Gru17].

Enhancement of the SWs amplitude and lateral GH shifts values are of main interest in our investigation. With this purpose we can select the most favorable parameters of the second ferromagnet. Hence, obtaining of the reflectance, transmittance and GH shifts extreme values can be treated as the multivariable optimization problem. Finding the Hessian matrix for the function of two variables and using Sylvester's criterion enables us to find the maximum and minimum values of the reflectance, transmittance and GH shifts.

### III. RESULTS

Let us analyze the results obtained for SWs of frequency 100 GHz in the medium composed of two semi-infinite ferromagnetic materials saturated by the static magnetic field of value 15 kOe oriented parallel to the $z$ axis (schematically presented in Fig. 1). One of them (FM-1) has in all calculations assumed magnetic parameters of Py (see magnetic parameters in Table 1). The magnetic parameters of the second ferromagnet (FM-2) differs. FM-1 and FM-2 are separated by the 2 nm narrow interface region with the same magnetic parameters as Py with one exception of anisotropy value that is varied. We consider here only SWs incident from Py under the angle of incidence 60°.

In Fig. 2 are presented dependencies of the GH shift in reflection (Fig. 2 (a)), the reflectance (Fig. 2(b)) and transmittance (Fig. 2(c)) on the saturation magnetization and exchange constant of the FM-2. We have assumed only physical range of $M_2$ and $A_2$ values in the FM-2, i.e. $M_2 \in [0, 2\times10^3]$ Gs and $A_2 \in [0, 3\times10^{-6}]$ erg/cm. The interface region is assumed to have Py saturation magnetization and exchange constant and non-zero interfacial anisotropy, $K_{12} = 2.5\times10^6$ erg/cm$^3$. It is visible that within that range of parameters all those functions changes significantly. Reflectance and transmittance takes broad part of the possible values. The GH shift in reflection can be either positive or negative, with the values up to 70 nm. Interestingly, maximal value of the GH shift in Fig. 2(a) corresponds to the values of saturation magnetization and exchange constant near those of Co. This maximal GH shift appears when the reflectance (transmittance) reach minimum (maximum). We see, that the increase of the exchange constant of the second medium will further increase the GH shift, but we can find an optimal value of the saturation magnetization which maximize the GH shift.

In order to precisely analyze SWs reflection and refraction at the interface between different ferromagnets we will consider in details the FM-2 for magnetic parameters of the most typical magnetic materials, i.e., Py, Co, Fe and Ni (see material parameters in the Table 1). We will analyze how SWs refraction and reflection changes while the interfacial anisotropy $K_{12}$ is varied within the physically realistic range of values, $K_{12} \in [-4.5\times10^6, 4.5\times10^6]$ erg/cm$^3$.[Vaz08] In Fig. 3 are presented results of the GH shift in reflection and transmission, the reflectance and transmittance in dependence on the anisotropy of the interfacial region. In the case of Py on both sides of the interface the GH shifts for reflected and refracted SWs are equal to each other, the case studied in Ref. [Gru17]. However, if we consider two different materials, $\Delta_r$ differs from $\Delta_t$. Moreover, that difference in GH shifts between the two different FM-2 materials is not only quantitative, but also qualitative. For the interfaces Py/Py (Fig. 3(a)), Py/Fe (Fig. 3(b)) and Py/Ni (Fig. 3(c)) there is monotonous change of the GH shift with increasing anisotropy constant at the interface. Also in these cases the shifts in reflection and transmission behaves similarly. The case of Py/Co is different (see Fig. 3(g)). There, for the values of $K_{12}$ near zero, $\Delta_r$ has positive slope whereas $\Delta_t$ has negative slope. In the considered range of $K_{12}$ the GH shift in reflection and transmission have opposite signs, reaching maximal absolute values (up to 100 nm) in the reflection for anisotropy equal ca. $10^6$ erg/cm$^3$. For the Py/Co the variation of the R and T in dependence on anisotropy at interface is most significant among considered materials.

## IV. Conclusion

We have performed extended analytical study of the SWs reflection and refraction at the interface between two semi-infinite ferromagnetic materials separated by the ultra-narrow interface region of the width much smaller than the wavelength of the SWs. We have focused on the purely exchange SWs incident from the Py to other ferromagnetic material to study the GH effect for SWs in reflection and transmission. We have shown that with the change of the second material, the significant change in the GH shift can be observed. We report, that for the interface Py/Co, a big GH shift of value up to 100 nm in reflection, which is around 5 times longer than the SW wavelengths in Py, can be observed. Presented results demonstrate the possibility for the SWs steering at subwavelength distances which can be the first step in designing magnonic metasurfaces.

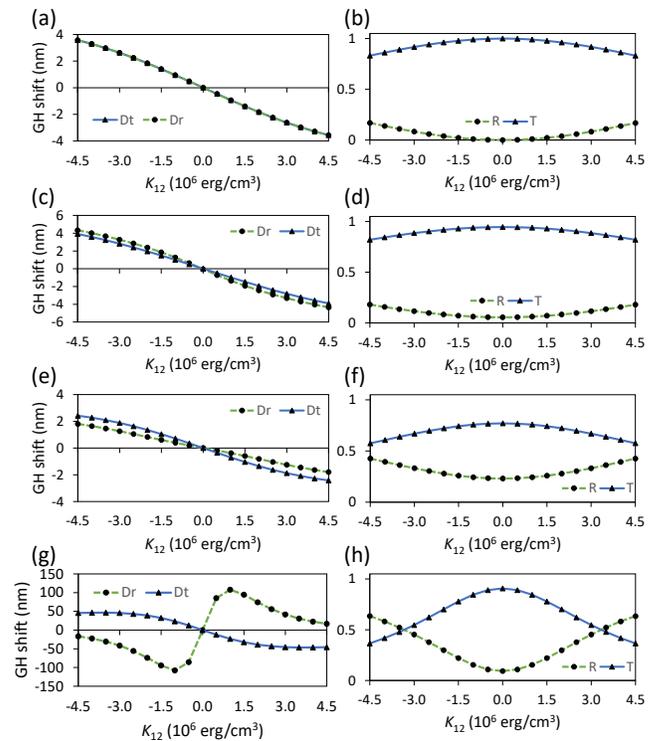

Fig. 3. Results of the analytical model for (a) and (b) Py/Py; (c) and (d) Py/Fe; (e) and (f) Py/Ni; (g) and (h) Py/Co interfaces. In (a), (c), (e) and (g) are presented GH shifts dependencies for reflected (Dr) and transmitted (Dt) SWs as a function on $K_{12}$. In (b), (d), (f) and (g) are presented reflectance and transmittance dependencies on value of $K_{12}$.

## Acknowledgments

This research has received founding from National Science Centre of Poland project No. UMO-2012/07/E/ST3/00538 and from the European Union Horizon2020 research and innovation programme under the Marie Sklodowska-Curie grant agreement No. 644348 (MagIC). The computations were partially performed at Poznan Supercomputing and Networking Center (grant No 209).